\documentclass[sigconf, screen, nonacm]{acmart}

\setcopyright{none}

\usepackage{amsmath}
\usepackage{graphicx}
\usepackage{booktabs,multirow}
\usepackage{enumitem}

\begin{document}


\title[FOREST: Heterogeneous Memory Coordination]{FOREST: Coordinating Tree-based HBF, RRAM, and HBM for Efficient Speculative LLM Inference}

\author{Hao Ding}
\affiliation{%
  \institution{Unigroup Guoxin Microelectronics Co., Ltd.}
  \country{China}
}

\author{Zheng Shi}
\affiliation{%
  \institution{School of Integrated Circuits, Peking University}
  \country{China}
}

\author{Dongxue Zhao}
\affiliation{%
  \institution{School of Integrated Circuits, Peking University}
  \country{China}
}

\author{Huiying Bian}
\affiliation{%
  \institution{National Space Science Center, Chinese Academy of Sciences}
  \country{China}
}

\author{Meng Li}
\affiliation{%
  \institution{School of Integrated Circuits, Peking University}
  \country{China}
}
\affiliation{%
  \institution{School of Computer Science, Peking University}
  \country{China}
}

\author{Xu Zheng}
\affiliation{%
  \institution{State Key Laboratory of Fabrication Technologies for Integrated Circuits, Institute of Microelectronics of Chinese Academy of Sciences}
  \country{China}
}

\author{Kun Wang}
\affiliation{%
  \institution{Unigroup Guoxin Microelectronics Co., Ltd.}
  \country{China}
}

\author{Nanjian Wu}
\affiliation{%
  \institution{Unigroup Guoxin Microelectronics Co., Ltd.}
  \country{China}
}

\author{Jie Chen}
\affiliation{%
  \institution{Unigroup Guoxin Microelectronics Co., Ltd.}
  \country{China}
}

\author{Yimao Cai}
\affiliation{%
  \institution{School of Integrated Circuits, Peking University}
  \country{China}
}
\renewcommand{\shortauthors}{H. Ding et al.}

\makeatletter
\renewcommand{\@mkauthors@iii}{%
  \global\setbox\mktitle@bx=\vbox{%
    \unvbox\mktitle@bx
    \vspace{0.65em}
    \centering

    {\large
      Hao Ding\textsuperscript{1},
      Zheng Shi\textsuperscript{2},
      Dongxue Zhao\textsuperscript{2},
      Huiying Bian\textsuperscript{3},
      Meng Li\textsuperscript{2, 5},
      Xu Zheng\textsuperscript{4},
      Kun Wang\textsuperscript{1},\\[0.15em]
      Nanjian Wu\textsuperscript{1},
      Jie Chen\textsuperscript{1},
      Yimao Cai\textsuperscript{2}\par
    }

    \vspace{0.45em}

    {\normalsize
      \textsuperscript{1}Unigroup Guoxin
      Microelectronics Co., Ltd., Beijing, China. 
      \textsuperscript{2}School of Integrated
      Circuits, Peking University, Beijing, China. 
      \textsuperscript{3}National Space Science
      Center, Chinese Academy of Sciences, Beijing, China. 
      \textsuperscript{4}
      Institute of Microelectronics of
      Chinese Academy of Sciences, Beijing, China. 
      \textsuperscript{5}School of Computer Science,
      Peking University, Beijing, China\par
    }
    
\vspace{0.3em}

{\normalsize\rmfamily
  Email:
  \{dingh,wunj\}@gosinoic.com,
  \{shizheng,zhaodongxue,meng.li,caiyimao\}@pku.edu.cn,
  bianhuiying@mirslab.cn,
  zhengxu2018@ime.ac.cn,
  wk1204@hotmail.com,
  chenjie@unigroup.com.cn\par
}

    \vspace{0.7em}
  }%
}


\makeatother

\begin{abstract}
Speculative decoding (SD) accelerates large language model (LLM) inference by combining lightweight token drafting with parallel target-model verification. However, its execution stages impose fundamentally different resource demands: target verification requires high-throughput access to large weight sets, sequential drafting repeatedly executes a compact model under tight latency constraints, and long-context attention processes dynamically growing key-value (KV) caches. Conventional GPU-centric SD systems serve these distinct workloads through a shared HBM-centric memory hierarchy and general-purpose compute resources, creating mismatches between memory capacity, bandwidth, and computational demands. These mismatches constrain deployable model--context configurations and limit the efficiency of both bandwidth-bound and latency-sensitive operations.

To address these challenges, we propose FOREST, a heterogeneous memory-centric architecture that coordinates HBF, 3D-stacked RRAM, and HBM memory--compute trees for efficient SD, featuring: (1) compute-aligned 3D RRAM architecture, combining ultra-wide, low-frequency reads with clock-decoupled interfaces and overlapped SRAM-DCIM weight refills for latency-sensitive drafting; (2) cross-tree numerical co-design, moving block-scaled Q/K/V preprocessing to HBF and RRAM root plates and enabling block-floating-point (BFP) attention with pipelined softmax reduction in the HBM tree; and (3) HBF-aware SD design-space exploration, revealing how bandwidth--compute provisioning and model scale determine performance bottlenecks and performance, energy, and cost trade-offs. Evaluated across six SD configurations spanning 32B--405B target models, FOREST achieves up to 5.5$\times$ end-to-end speedup, 11.46$\times$ higher decode throughput, and 16.5$\times$ higher decode energy efficiency over NVIDIA A800 GPU-based SD baselines. Under a modeled total cost of ownership, FOREST further delivers up to 6.73$\times$ higher token-generation cost efficiency.
\end{abstract}

\maketitle

\section{Introduction}
\label{sec:introduction}

Large language models (LLMs) have demonstrated remarkable capabilities across a wide range of applications, yet their growing model sizes and context lengths pose significant challenges to efficient inference\cite{DBLP:journals/corr/VaswaniSPUJGKP17}. Conventional autoregressive decoding generates tokens sequentially, resulting in substantial inference latency and repeated execution of expensive model computations. Speculative decoding (SD) addresses this bottleneck by employing a lightweight draft language model (DLM) to propose multiple candidate tokens, which are subsequently verified in parallel by a larger target language model (TLM)\cite{pmlr-v202-leviathan23a,chen2023acceleratinglargelanguagemodel}. By amortizing expensive target-model execution over multiple accepted tokens, SD can improve LLM inference throughput.

However, the drafting, verification, and attention stages of SD impose distinct memory and computational demands, particularly as context lengths increase. TLM verification processes multiple candidate tokens in parallel, demanding efficient access to large model weights and high computational throughput. In contrast, the lightweight DLM generates tokens sequentially and is invoked repeatedly, making low-latency execution and efficient access to its compact weight set critical\cite{10.1145/3620666.3651352}. Meanwhile, attention in both models must access and update their respective key-value (KV) caches, whose storage footprint and data movement costs grow with context length\cite{10.1145/3600006.3613165}. Consequently, SD combines capacity-intensive, bandwidth-intensive, and latency-sensitive workloads within a single inference system.

In conventional GPU-based SD deployments, the TLM, DLM, and their respective KV caches typically share an HBM-centric memory hierarchy and general-purpose compute fabric~\cite{Miao_2024}. This shared resource organization serves workloads with distinct capacity, bandwidth, and latency requirements through the same hardware substrate. As model sizes and context lengths increase, large weight footprints and growing KV caches place increasing pressure on memory capacity and bandwidth, limiting deployable model--context configurations or requiring costly memory expansion~\cite{10.1145/3600006.3613165,sheng2023flexgen}. Meanwhile, sequential drafting and bandwidth-bound operations often underutilize the available GPU computational resources~\cite{wang2026saddle}. Expanding memory capacity does not resolve the mismatch between data delivery and compute consumption, while increasing compute capacity cannot eliminate memory bandwidth bottlenecks. Efficient SD therefore requires memory and compute resources tailored to the execution characteristics of drafting, verification, and attention.

Prior studies have improved SD efficiency through algorithmic and architectural optimizations. EAGLE-3~\cite{li2025eagle3} enhances draft prediction, while SpecPIM~\cite{10.1145/3620666.3651352} and HybridSpec~\cite{huang2026hybridspec} explore heterogeneous hardware resources and execution dataflows. H$^3$~\cite{ha2026h3} further integrates HBM and high-bandwidth flash (HBF) to expand memory capacity for largely static LLM weights. Beyond memory capacity, sustained computing throughput requires matching memory access granularity and data delivery rates to the consumption patterns of compute arrays~\cite{wang2026flashaccel,oliveira2026flint}. Moreover, distributing projections and attention across different computing substrates requires coordinated numerical representations to manage exponent scaling, format conversion, and intermediate data movement~\cite{chen2026p3llm,wang2026harmonia}.

Recent advances in three-dimensional (3D) memory integration have expanded the design space of high-bandwidth memory systems beyond conventional DRAM-based HBM to emerging nonvolatile technologies, such as HBF and resistive random-access memory (RRAM)~\cite{park2026hbmhbf,dong2026reramstacked}. HBF combines the high storage density of NAND flash with massively parallel read interfaces, offering the potential for substantially greater capacity and lower cost per bit when storing large, predominantly static TLM weights~\cite{ha2026h3,wang2026flashaccel}.

In contrast, sequential drafting repeatedly accesses a compact, largely static DLM weight set, making weight-access energy and latency critical. Although HBM offers greater capacity and aggregate bandwidth, serving these repeated weight accesses through HBM would also consume memory resources needed for dynamically updated KV caches. 3D-stacked RRAM offers fine-grained, low-latency, and energy-efficient parallel weight access~\cite{dong2026reramstacked}, providing a dedicated path that can reduce recurring weight-access energy while isolating DLM weight traffic from HBM-resident KV accesses. Together, these complementary memory characteristics motivate heterogeneous memory-centric architectures that coordinate storage, data delivery, and computation across distinct SD workloads~\cite{10.1145/3600006.3613165,park2024attacc}.

\begin{figure}[t]
    \centering
    \includegraphics[width=0.90\columnwidth]{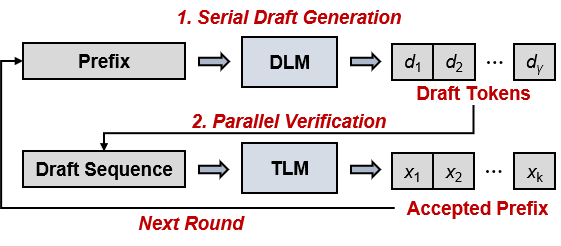}
    \caption{Overview of speculative decoding with sequential drafting and parallel verification.}
    \label{fig:sd_mechanism}
\end{figure}

\begin{table}[t]
\centering
\caption{Representative specifications of heterogeneous memory technologies.}
\label{tab:memory_comparison}
\scriptsize
\setlength{\tabcolsep}{2pt}
\renewcommand{\arraystretch}{1.15}

\resizebox{\columnwidth}{!}{%
\begin{tabular}{@{}lccccc@{}}
\toprule
Memory & Capacity & Read BW & Read Lat. & Read Energy  \\
       &          &         &           & (pJ/bit)    \\
\midrule
HBM~\cite{micronhbm2e,moon2023hbm,micron2026hbm4} &
\shortstack{2E: 16 GB\\4: 36 GB} &
\shortstack{410 GB/s/stack\\$>2.8$ TB/s/stack} &
$\sim$100 ns &
\shortstack{2E: 4.3\\4: N/R}  \\
\midrule
HBF~\cite{sandisk2025hbf, yu2026hbf} &
512 GB &
1.6 TB/s/stack &
$\sim$20 $\mu$s &
$\sim$6.8  \\
\midrule
RRAM~\cite{yang2024rram128m, crossbar_reram, dong2026reramstacked} &
\shortstack{32--100 MB/die} &
$\leq$1 GB/s/die &
15 ns &
0.36--0.44  \\
\bottomrule
\end{tabular}%
}

\end{table}

\begin{figure}[t]
    \centering
    \includegraphics[width=\columnwidth]{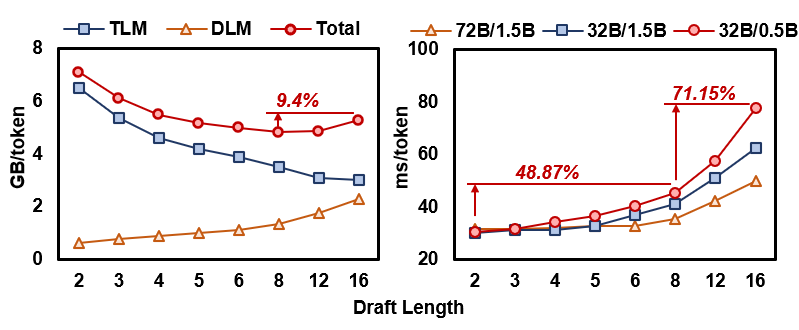}
    \caption{TLM--DLM resource trade-offs across draft lengths: (a) trace-derived logical W4-equivalent weight-read demand; (b) measured A800 GPU decode latency.}
    \label{fig:sd_characterization}
\end{figure}

Building on these opportunities, we propose \textbf{FOREST}, a heterogeneous memory-centric architecture that integrates HBF, HBM, and two 3D-stacked RRAM trees through workload-specialized SRAM-based digital computing-in-memory (DCIM) root plates. The HBF tree provides large-scale weight storage and parallel computation for TLM verification, while the RRAM trees support low-latency, repeated DLM execution. The HBM tree maintains dynamically updated KV caches and performs long-context attention through specialized K/V computation and hierarchical reduction. FOREST further coordinates these computing substrates by performing block-scaled Q/K/V preprocessing at the HBF and RRAM roots, enabling block-floating-point (BFP) based attention execution in the HBM tree. The main contributions of this work are as follows:

\begin{itemize}
    \item \textbf{Compute-aligned 3D RRAM architecture.}
    We design a 3D-stacked RRAM drafting architecture that combines ultra-wide, low-frequency sensing with clock-decoupled data delivery and overlapped SRAM-DCIM weight refills. This organization matches RRAM readout rates to compute-side consumption, sustaining low-latency DLM execution.

    \item \textbf{Cross-tree numerical and attention co-design.}
    We distribute block-scaled Q/K/V preprocessing across the HBF and RRAM roots and develop an HBM-resident attention architecture with asymmetric K/V datapaths, 16-D BFP computation, local softmax, and hierarchical global reduction. This design coordinates numerical processing across memory trees while reducing centralized attention data movement.

    \item \textbf{SD-aware heterogeneous execution and provisioning.}
    We coordinate TLM verification, DLM drafting, and shared HBM attention through phase-aware execution and memory--compute resource allocation. We further characterize how HBF bandwidth and compute provisioning interact with model scale and speculative execution, revealing performance--energy--cost trade-offs across heterogeneous SD configurations.
\end{itemize}

Our evaluation across six target--draft configurations spanning 32B--405B target models demonstrates that FOREST achieves up to 5.5$\times$ end-to-end speedup, 11.46$\times$ higher decode throughput, and 16.5$\times$ higher decode energy efficiency compared with NVIDIA A800 GPU-based SD baselines. Under a modeled total cost of ownership, FOREST further delivers up to 6.73$\times$ higher token-generation cost efficiency.

\begin{figure*}[t]
    \centering
    \includegraphics[width=\textwidth]{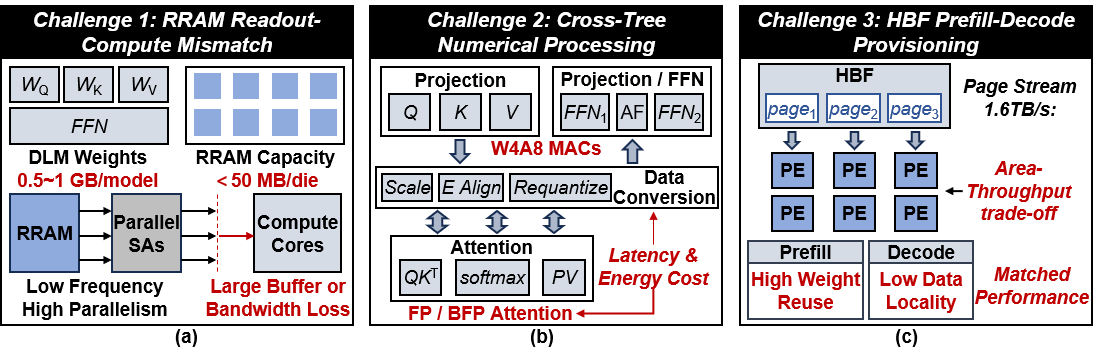}
    \caption{Architectural challenges: (a) RRAM sensing-to-compute bandwidth mismatch; (b) cross-tree numerical processing overhead; (c) HBF bandwidth--compute trade-offs.}
    \label{fig:forest_challenges}
\end{figure*}
\section{Background}

\subsection{Speculative Decoding Execution Model}
\label{subsec:sd_execution}

As illustrated in Fig.~\ref{fig:sd_mechanism}, SD alternates between sequential drafting and parallel verification~\cite{sun2025blockverification}. A DLM autoregressively proposes up to $\gamma$ candidate tokens, which a TLM verifies in a single forward pass. Verification commits the accepted prefix and may append a correction or bonus token before the next round~\cite{kumar2026ssd}. Consequently, the effective progress of SD depends on the number of committed tokens per iteration rather than the draft length alone~\cite{li2024eagle2,zhang2025svip}.

SD also maintains separate KV caches for the DLM and TLM~\cite{kim2026specllm}. During drafting, the DLM incrementally updates its KV cache, while the TLM computes candidate KV states during parallel verification~\cite{li2026specstream}. Since only the accepted prefix is committed, KV entries associated with rejected candidates must be invalidated to prevent their reuse in subsequent rounds~\cite{kim2026specllm,li2026specstream}. These KV updates and validity-management operations introduce additional overhead alongside drafting and verification, contributing to the execution cost per committed token~\cite{kim2026specllm}.

\subsection{Heterogeneous 3D Memory Integration}
\label{subsec:3d_memory}

3D die stacking integrates multiple memory dies with a logic base through high-density vertical interconnects, such as through-silicon vias (TSVs) and hybrid bonding~\cite{lee2025hbm3d,huang2025hbminterconnect}. HBM stacks DRAM dies to provide high bandwidth and efficient random read/write access, although capacity expansion remains costly~\cite{samsung2026hbm4,ou2025opticalhbm}. In contrast, HBF leverages dense 3D NAND to provide substantially larger storage capacity with HBM-class aggregate read bandwidth~\cite{ocp2026hbf,skhynix2026hbf}. Its page-oriented access and relatively long sensing latency require sufficient parallelism and buffering to sustain data delivery~\cite{jonnalagadda2026splash}.

RRAM offers nonvolatile storage with low-latency reads and highly parallel sensing capabilities~\cite{lin2024nearram}. Unlike monolithic 3D (M3D) integration, which fabricates additional memory layers directly above an underlying device layer~\cite{xie2023m3d,qiu2025m3d}, die-stacked RRAM integrates separately fabricated memory and logic dies through vertical interconnects. FOREST adopts this organization, building on recent ReRAM-on-logic accelerators~\cite{dong2026reramstacked}, to combine independently fabricated memory arrays and compute logic with parallel readout across stacked dies. Table~\ref{tab:memory_comparison} summarizes the representative characteristics and technology maturity of these memory technologies.

\section{Challenges and Motivation}
\label{sec:motivation}

\subsection{Resource Trade-offs in SD}
\label{subsec:hierarchical_demands}
SD couples large-scale target-model verification with latency-sensitive sequential drafting~\cite{yan2025decoding}. Increasing the draft length $\gamma$ amortizes TLM weight accesses over more committed tokens but introduces additional DLM execution steps~\cite{huang2024specdecpp}. Fig.~\ref{fig:sd_characterization}(a) quantifies their logical weight-read demands using measured SD traces and W4-equivalent model sizes. For Qwen2.5-32B/1.5B, increasing $\gamma$ from 2 to 8 reduces TLM demand from 6.48 to 3.49 GB/token while increasing DLM demand from 0.61 to 1.32 GB/token. The combined demand decreases by 32.2\% to 4.81 GB/token, but rises again at longer draft lengths, revealing opposing weight-delivery pressures between verification and drafting.

However, lower logical weight-read demand does not necessarily translate into lower execution latency~\cite{sadhukhan2025magicdec}. As shown in Fig.~\ref{fig:sd_characterization}(b), increasing $\gamma$ from 2 to 8 raises the measured GPU decode latency from 30.15 to 41.10 ms/token for Qwen2.5-32B/1.5B, even as the combined weight-read demand decreases by 32.2\%. Moreover, at $\gamma=8$, the smaller 0.5B DLM requires 27.6\% more verification rounds than the 1.5B DLM and exhibits 10.4\% higher decode latency under the same 32B target. These results demonstrate that SD execution efficiency depends on the interplay among weight-delivery bandwidth, sequential drafting costs, and acceptance-driven verification frequency, rather than model size or memory traffic alone.

\begin{figure*}[t]
    \centering
    \includegraphics[width=0.8\textwidth]{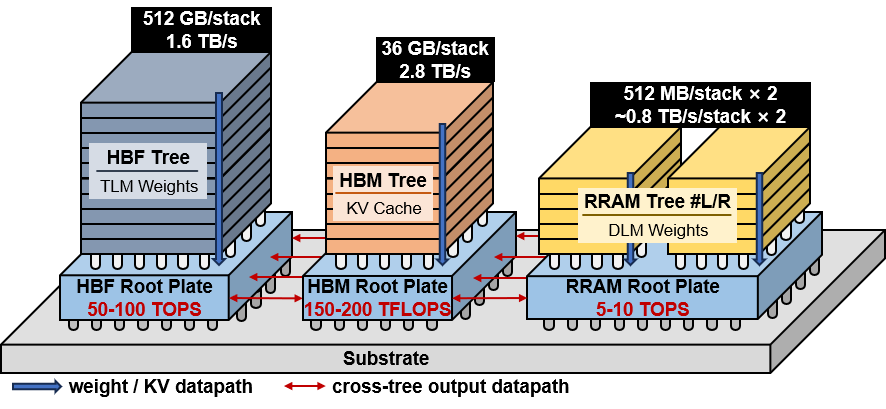}
    \caption{Overall organization of FOREST with heterogeneous memory trees and workload-specialized compute root plates.}
    \label{fig:forest_overview}
\end{figure*}

\subsection{Challenges}
\label{subsec:heterogeneous_challenges}
SD exposes a structural mismatch between the aggregate capabilities of heterogeneous memories and the requirements of coordinated inference execution. Efficiently combining HBF, RRAM, and HBM requires matching memory delivery to compute consumption, coordinating numerical processing across memory--compute roots, and balancing bandwidth and computation across execution phases. As illustrated in Fig.~\ref{fig:forest_challenges}, these requirements give rise to three architectural challenges that motivate the design of FOREST.

First, dedicated RRAM storage can provide energy-efficient access to repeatedly used DLM weights while isolating weight traffic from HBM-resident KV accesses. However, exploiting RRAM's parallel readout for sustained drafting introduces a mismatch between memory sensing and compute consumption~\cite{dong2026reramstacked,chi2016prime}. RRAM arrays operate at relatively low sensing frequencies, producing wide data bursts that must traverse die-to-logic interfaces before reaching SRAM-DCIM compute units. Mismatches among sensing, transmission, and buffer-refill rates can stall computation or require substantial intermediate buffering~\cite{ankit2019puma}. 

Second, separating weight computation from KV-resident attention creates numerical processing demands across memory--compute trees. Low-precision projection datapaths generate Q/K/V activations, while attention requires sufficient dynamic range for score accumulation, softmax normalization, and value aggregation~\cite{fang2025anda,stevens2021softermax}. Performing exponent extraction, mantissa scaling, and format conversion entirely within the attention root concentrates peripheral computation and buffering~\cite{chen2026p3llm}. This motivates distributing block-scaled preprocessing between the Q/K/V-producing roots and the HBM attention datapath while maintaining the numerical range required for attention.

Third, high aggregate HBF bandwidth does not necessarily translate into efficient TLM execution~\cite{wang2026flashaccel,oliveira2026flint}. Target verification benefits from weight reuse across candidate positions, whereas long-prompt prefill exposes greater computational parallelism and can become compute-bound even with sufficient memory bandwidth~\cite{patel2024splitwise,zhong2024distserve}. Conversely, additional compute capacity offers limited benefits when page-oriented weight delivery becomes the bottleneck~\cite{williams2009roofline,oliveira2026flint}. HBF resource provisioning must therefore account for the changing bandwidth and compute bottlenecks across prefill and verification.

\subsection{Motivation}

These challenges call for coordinated memory--compute design rather than independent optimization of memory bandwidth or compute throughput. FOREST addresses three coupled requirements: rate-matched RRAM weight delivery for drafting, producer-side numerical preprocessing coordinated with HBM-resident attention, and HBF bandwidth--compute provisioning across prefill and verification. Together with phase-aware execution and KV-state management, these mechanisms target the end-to-end execution cost per committed token.

\section{Hardware Design Details}
\label{sec:overview}

\subsection{Architecture Overview}
\label{subsec:architecture_overview}

\begin{figure*}[t]
    \centering
    \includegraphics[width=0.88\textwidth]{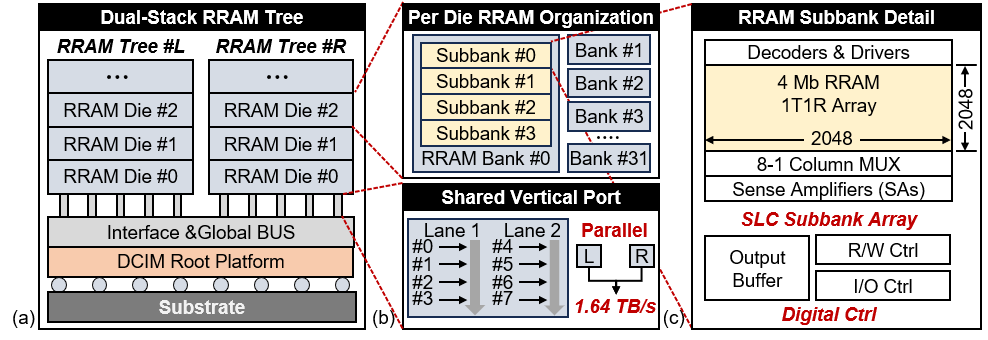}
    \caption{Die-stacked RRAM tree organization with hierarchical memory banks and shared vertical read interfaces.}
    \label{fig:rram_tree}
\end{figure*}

Fig.~\ref{fig:forest_overview} illustrates FOREST's heterogeneous memory--compute architecture, comprising an HBF tree, two RRAM trees, and an HBM tree, each coupled with a workload-specialized SRAM-based DCIM root plate. The HBF tree stores large-scale TLM weights and supports throughput-oriented prefill and parallel verification, while the RRAM trees provide repeated weight delivery for latency-sensitive DLM drafting. The HBM tree maintains separate, dynamically updated KV caches for both models and executes attention using asymmetric K/V datapaths, local softmax, and hierarchical global reduction. Its attention root is time-shared between TLM verification and DLM drafting, while the two models retain independent KV histories.

FOREST keeps model-weight accesses and historical KV accesses local to their respective memory trees, exchanging intermediate activations across the compute roots. The HBF and RRAM roots perform weight-dependent computation and generate block-scaled Q/K/V activations, which are forwarded with their exponent metadata to the HBM attention root. The HBM root stores newly generated KV states and returns attention outputs to the corresponding weight-compute roots for subsequent layer computation. This organization coordinates projection and attention execution across heterogeneous memories without transferring full model weights or KV histories between trees.

\subsection{Dual-Channel RRAM Tree}
\label{sec:RRAM organization}

FOREST employs two eight-die RRAM stacks sharing an SRAM-DCIM root platform for latency-sensitive DLM drafting, as shown in Fig.~\ref{fig:rram_tree}. Each 64-MiB die contains 32 banks divided into four subbanks each, providing 1 GiB of aggregate weight capacity across the 16 dies. Bank-level parallel sensing produces a 32,768-bit readout per active die at 100 MHz, while each stack connects to 128 SRAM-DCIM processing elements (PEs) through shared vertical interfaces.

\begin{figure}[t]
    \centering
    \includegraphics[width=0.95\columnwidth]{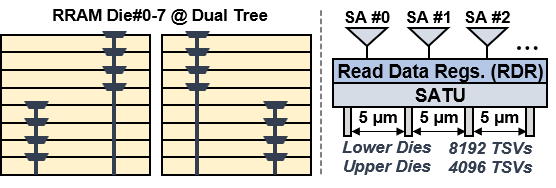}
    \caption{Half-stack shared vertical interconnect and die-local RRAM readout interface.}
    \label{fig:rram_tsv}
\end{figure}

\begin{figure}[t]
    \centering
    \includegraphics[width=\columnwidth]{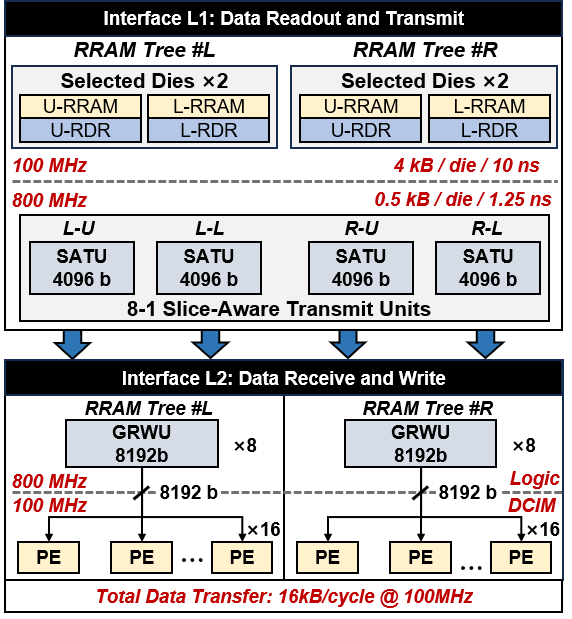}
    \caption{Two-level RRAM interface for parallel readout, high-speed transmission, and SRAM-DCIM weight delivery.}
    \label{fig:rram_interface}
\end{figure}

To match RRAM sensing with compute-side weight consumption, FOREST employs a two-level clock-decoupled interface, illustrated in Figs.~\ref{fig:rram_tsv} and~\ref{fig:rram_interface}. Die-local read data registers (RDRs) capture each 32,768-bit sensing result, while an 8:1 slice-aware transmit unit (SATU) serializes the readout into eight 4,096-bit transfers at 800 MHz. Two half-stack lanes operate concurrently, delivering 819.2 GB/s per stack and 1.64 TB/s across both stacks. At the root platform, grouped receive and write units (GRWUs) assemble eight 8,192-bit beats into a 65,536-bit frame. Grouped interface units (GIUs) then distribute each frame across eight groups of 16 PEs, supplying 512 bits per PE at 100 MHz.

Each PE employs two 4-KiB ping-pong weight-buffer partitions to overlap computation with the next weight-tile refill. While one partition supplies the active computation, the other receives incoming weights. The compute path reads 2,048 bits per cycle through 16 parallel 128-bit SRAM banks, whereas the refill path writes four 128-bit banks per cycle, providing 512 bits of incoming weight data. At this rate, a complete 4-KiB partition refill takes 64 cycles. The two partitions alternate after their respective computation and refill operations complete, allowing weight delivery to overlap with ongoing DCIM execution.

\begin{figure}[t]
    \centering
    \includegraphics[width=0.75\columnwidth]{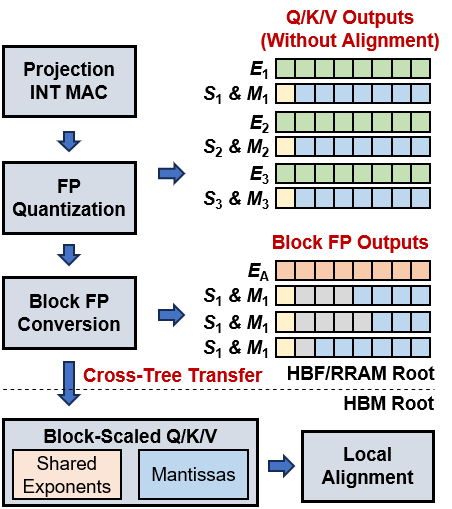}
    \caption{Cross-tree numerical preprocessing for block-scaled attention computation.}
    \label{fig:cross_tree_numerical}
\end{figure}

\subsection{HBF Weight Delivery and Cross-Tree Numerical Dataflow}
\label{sec:overview_dataflow}

FOREST employs page-aligned weight delivery to match HBF accesses with SRAM-DCIM computation. Each 4-KiB HBF page maps directly to a PE-local weight tile. Across 16 memory channels, page buffers and row formatters reorganize 256-byte transfers at 400 MHz into 512-byte row writes at 200 MHz, preserving the channel data rate while matching the DCIM weight-buffer interface. Channel-aware loaders distribute weight tiles across 40 PEs per channel. Once loaded, each tile is reused across verification candidates, amortizing HBF page accesses over multiple computations.

FOREST further distributes numerical processing across the HBF, RRAM, and HBM roots, as illustrated in Fig.~\ref{fig:cross_tree_numerical}. Following integer projections, the HBF and RRAM roots perform 16-D block-scaled Q/K/V preprocessing, including shared-exponent extraction, mantissa alignment, and quantization. The resulting block-floating-point (BFP) mantissas and exponent metadata are forwarded to the HBM attention root. Its K-oriented datapath performs exponent-aware QK$^\mathrm{T}$ computation and score accumulation, followed by local softmax and V-oriented weighted-value accumulation. A hierarchical reduction network then merges partial attention results across processing units. By preparing BFP operands at the producing roots, FOREST distributes numerical conversion across memory trees and enables direct consumption of block-scaled activations by the HBM attention datapath.

\subsection{SRAM DCIM Root Plate}
\label{sec:overview_organization}

\begin{figure}[t]
    \centering
    \includegraphics[width=\columnwidth]{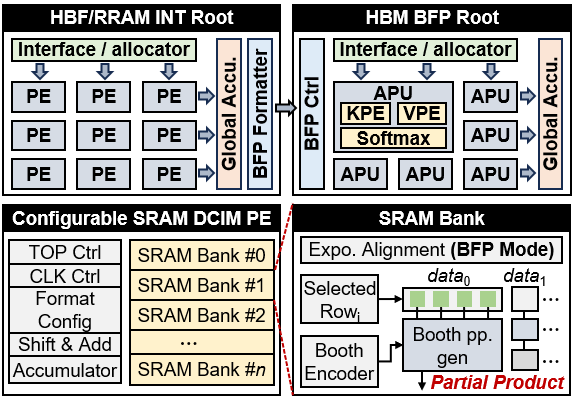}
    \caption{Workload-specialized root-plate organizations and the configurable SRAM-DCIM PE substrate.}
    \label{fig:root_pe_organization}
\end{figure}

FOREST builds workload-specialized root plates on a configurable SRAM-based DCIM substrate, as illustrated in Fig.~\ref{fig:root_pe_organization}. The HBF and RRAM roots employ integer-oriented PEs for W4A8 projection and feed-forward computation, with 40 logical PE slots per HBF channel and 128 PEs per RRAM tree. The HBM root integrates 512 attention processing units (APUs), each comprising K- and V-oriented processing elements and local softmax logic. These root plates share a banked SRAM-DCIM organization while employing operation-specific arithmetic and control. Integer projection and BFP attention are implemented through specialized datapaths rather than full floating-point MAC units in individual SRAM banks.

Each HBM APU integrates a K-PE and a V-PE with asymmetric computation and bank mappings. Both employ 32 banks, eight rows per bank, and 16 lanes per row, but specialize in different attention operations. 
The K-PE computes query-key dot products of up to 128 dimensions, organized into eight 16-dimensional BFP groups, with exponent-scaled partial sums aligned to produce attention scores. Local softmax converts these scores into 8-bit attention coefficients, which are consumed by the V-PE for weighted-value accumulation. The V-PE partitions its 32 banks into four groups of eight and employs four-phase Booth-based multiplication to sequentially update 128 running accumulators while maintaining BFP scaling information. This asymmetric organization supports score reduction and value aggregation through distinct computation schedules on the same SRAM-DCIM substrate.

Each APU generates a local maximum score, an exponential normalization sum, and a weighted-value numerator. A hierarchical global merge tree consisting of 511 binary merge nodes recursively combines the partial results from all 512 APUs. At each merge node, the local maxima establish a common normalization reference, while the exponential sums and BFP weighted-value numerators are rescaled and accumulated accordingly. The root produces the final attention output by normalizing the merged weighted-value numerator with the global exponential sum. This hierarchical organization performs distributed softmax reduction without transferring intermediate attention scores and value vectors to a centralized computation unit.

\begin{figure*}[t]
    \centering
    \includegraphics[width=0.86\textwidth]{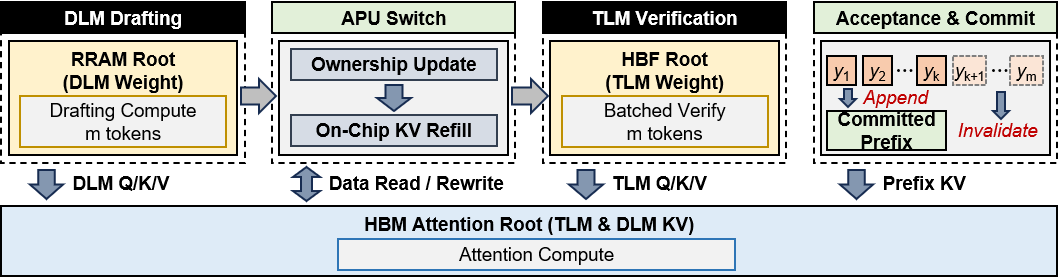}
    \caption{Phase-aware speculative decoding execution in FOREST with shared HBM attention-root reuse.}
    \label{fig:sd_execution}
\end{figure*}

\section{Algorithm Mapping and Execution}
\label{sec:attention}
\subsection{Model and State Mapping}
\label{sec:attention_chunks}

FOREST maintains independent TLM and DLM KV caches in separate HBM regions while time-multiplexing a shared 512-APU attention root. Rather than statically partitioning the array between models, FOREST assigns all 512 APUs to the active model during each execution phase. The K- and V-PEs within each APU switch ownership together, ensuring that their resident KV tiles belong to the same model. At phase transitions, ownership metadata invalidate the outgoing model's PE-resident working set, while the incoming model's KV tiles are refilled from HBM on demand. The corresponding BFP mantissas and shared-exponent metadata follow the same ownership and refill rules, leaving the model-specific KV histories intact in HBM.

During speculative execution, the DLM incrementally generates tentative KV entries, while the TLM computes candidate KV states during parallel verification. After the accepted prefix is determined, model-specific validity boundaries retain committed entries and logically invalidate rejected suffixes without physically clearing the memory. Missing KV states for newly committed correction or bonus tokens are generated when the corresponding model next processes those tokens. Together, phase-level APU ownership switching and model-specific KV validity management enable the two models to share the attention hardware without transferring their complete KV histories between memory regions.

\subsection{Phase-Aware Execution and Pipeline Coordination}
\label{subsec:phase_aware_execution}

Fig.~\ref{fig:sd_execution} illustrates FOREST's phase-aware execution schedule, which alternates DLM drafting and TLM verification on the shared HBM attention root. During drafting, the RRAM root sequentially generates up to $\gamma$ candidate tokens, invoking the HBM attention root with DLM-specific KV states. Once outstanding attention operations complete, the controller switches all 512 APUs to TLM ownership and refills the required KV working tiles on demand. The HBF root then verifies the candidates in parallel, reusing streamed weight tiles across candidate positions while the HBM root performs target-model attention. After verification, the controller commits the accepted prefix, updates model-specific KV validity boundaries, and returns the APUs to DLM ownership for the next speculative round.

FOREST overlaps memory delivery and computation through tile-granular pipelining within each execution phase. During drafting, RRAM weight refills overlap computation on active SRAM-DCIM buffers, while verification reuses HBF weight tiles across candidate positions. As block-scaled Q/K/V activations arrive from the producing roots, the HBM attention root processes them through K-side score generation, local online softmax, V-side weighted accumulation, and hierarchical global reduction. At phase boundaries, the controller overlaps KV working-set refills with independent producer-root operations whenever their data dependencies allow, reducing exposed phase-transition latency.

\section{Experimental Setup}
\label{sec:experimental_setup}

\subsection{Baselines and Workloads}
\label{subsec:baselines_workloads}

We evaluate FOREST across six target--draft configurations: Qwen2.5-32B with 0.5B, 1.5B, and 3B autoregressive (AR) drafts; Qwen2.5-72B with a 1.5B draft; GLM-4.5-Air (106B) with native multi-token prediction (MTP); and Llama 3.1-405B with a Llama 3.2-1B draft~\cite{qwen2025qwen25technicalreport,zeng2025glm45,llama2024herd}. GPU baselines use NVIDIA A800 80-GB accelerators: one GPU for Qwen models, two for GLM-4.5-Air with tensor parallelism (TP) and expert parallelism (EP), and four for Llama 3.1-405B with four-way TP.

We select SD prompts spanning four workloads---general text, reasoning, code, and dialogue history---with two 8K-token prompts per workload. We evaluate five output-token limits (128, 512, 1024, 2048, and 4096) and eight speculation lengths, $\gamma\in\{2,3,4,5,6,8,12,16\}$, together with target-only decoding. All experiments use batch size one and greedy decoding.

\subsection{Evaluation Methodology}
\label{subsec:evaluation_methodology}

We adopt trace-driven evaluation using measured GPU SD executions. For each model pair and draft length $\gamma$, we record draft proposals, acceptance decisions, verification rounds, and output tokens, which are replayed in FOREST's architectural model. GPU baselines are measured on NVIDIA A800 accelerators, excluding model initialization. Request latency includes both prefill and decoding, while GPU energy is collected through NVIDIA Management Library (NVML) counters and aggregated across devices for multi-GPU configurations~\cite{nvidiaNVML}. Comparisons are conducted at the system level using the GPU configurations specified in Section~\ref{subsec:baselines_workloads}. We report time to first token (TTFT), end-to-end latency, average power, and both end-to-end and decode-only throughput (tokens/s) and energy efficiency (tokens/J). All token-based metrics count committed output tokens. For optimized SD, each platform selects its draft length independently from the same eight-point sweep, while fixed-$\gamma=8$ comparisons provide a common configuration. GPU target-only inference serves as the non-speculative normalization reference.

FOREST's performance and energy are evaluated by combining trace replay with RTL verification, synthesis, and memory models. Icarus Verilog 13.0 and Yosys 0.56 support functional and structural verification, while Synopsys Design Compiler R-2020.09 provides standard-cell area, timing, and activity-based power estimates using a 22-nm IX22ULL library. SRAM costs are modeled with CACTI 7.0.3DD~\cite{balasubramonian2017cacti7}, and HBF, RRAM, and HBM are characterized by their bandwidth, latency, capacity, and access-energy parameters. The system simulator integrates these component characteristics with the replayed execution traces to estimate request-level latency and energy.

We evaluate total cost of ownership (TCO), including hardware acquisition, amortization, operation, and PUE-adjusted energy costs~\cite{greengrid2012pue}. The cost model includes memory stacks, compute logic, and heterogeneous integration, with hardware prices examined through sensitivity analysis. Cost efficiency is defined as
\begin{equation}
\mathrm{Tokens/\$} =
\frac{3600uT_{\mathrm{out}}}
{C_{\mathrm{hw}}/(8760L)+C_{\mathrm{op}}+C_{\mathrm{energy}}},
\label{eq:tokens_dollar}
\end{equation}
where $u$ is utilization, $T_{\mathrm{out}}$ is request-level output throughput including prefill and verification overheads, $C_{\mathrm{hw}}$ is hardware cost, $L$ is lifetime in years, and $C_{\mathrm{op}}$ and $C_{\mathrm{energy}}$ are hourly operating and energy costs. The reference TCO model assumes a four-year lifetime, a PUE of 1.3, an electricity cost of \$0.12/kWh, and hardware costs of \$20,000 per A800 GPU and \$22,000 for FOREST.

\begin{table}[t]
\centering
\caption{Hardware Configurations of the GPU Baseline and FOREST.}
\label{tab:hardware_config}

\normalsize
\setlength{\tabcolsep}{2pt}
\renewcommand{\arraystretch}{1.04}

\begin{tabular}{@{}p{0.38\columnwidth}p{0.59\columnwidth}@{}}
\toprule
\multicolumn{2}{c}{\textbf{Baseline: NVIDIA A800 GPU}} \\
\midrule
GPU Model & A800 80GB PCIe \\
GPU Count & 1 (Qwen); 2 (GLM-4.5-Air, TP+EP); 4 (Llama 405B, TP4) \\
Memory Capacity & 80 GB HBM2e/GPU \\
Memory Bandwidth & 1,935 GB/s/GPU \\
Peak Performance & 312 TFLOPS (BF16)/GPU \\
Interconnect & PCIe 4.0 $\times$16 \\
TDP & 300 W/GPU \\

\midrule
\multicolumn{2}{c}{\textbf{Proposed: FOREST System}} \\
Architecture & HBF T-tree + 2 RRAM D-trees + HBM KV-tree \\
Logic Technology & 22 nm (IX22ULL HD RVT) \\

\midrule
\multicolumn{2}{c}{\textbf{HBF Tree: TLM Verification}} \\
Memory Capacity & 512 GB/stack \\
Memory Bandwidth & 1.6 TB/s \\
Compute Organization & 640 W4A8 DCIM PEs \\
Weight Buffer & 4 KiB/PE \\
Peak Performance & 64 TOPS \\

\midrule
\multicolumn{2}{c}{\textbf{RRAM Trees: DLM Drafting}} \\
Memory Capacity & 1 GiB (2 stacks $\times$ 8 dies $\times$ 64 MiB) \\
Parallel Readout & 32,768 bits @ 100 MHz/die \\
Vertical Interface & 2 $\times$ 4,096 bits @ 800 MHz/tree \\
Memory Bandwidth & 819.2 GB/s/tree; 1.64 TB/s total \\
Root Configuration & 2 $\times$ 128 SRAM-DCIM PEs \\
Weight Buffer & 8 KiB/PE (2 $\times$ 4 KiB) \\
Peak Performance & 6.55 TOPS \\
Compute Precision & W4A8 \\

\midrule
\multicolumn{2}{c}{\textbf{HBM Tree: KV Attention}} \\
Memory Capacity & 36 GB/stack \\
Memory Bandwidth & 2.8 TB/s \\
Compute Organization & 512 APUs \\
Root Storage & 4.25 MiB total (K + V) \\
Compute & 192 TFLOPS-eq (K: 96 + V: 96) \\
Precision & BF16 KV; 16-D BFP internal \\

\bottomrule
\end{tabular}

\vspace{3pt}

\end{table}

\subsection{Hardware Configurations}
\label{subsec:forest_configuration}

Table~\ref{tab:hardware_config} summarizes the hardware configurations of both the NVIDIA A800 GPU baselines and FOREST. The GPU baseline includes device specifications and model-dependent GPU allocations, while FOREST specifies the memory capacities, bandwidths, compute organization, and arithmetic precision of its HBF, RRAM, and HBM trees. For GLM-4.5-Air, the native MTP layer serves as the drafter, with 75 routed experts mapped to RRAM and the remaining 53 to HBF under uniform top-8 expert routing.
\section{Evaluation}
\label{sec:evaluation}

\begin{figure*}[t]
    \centering
    \includegraphics[width=0.9\textwidth]{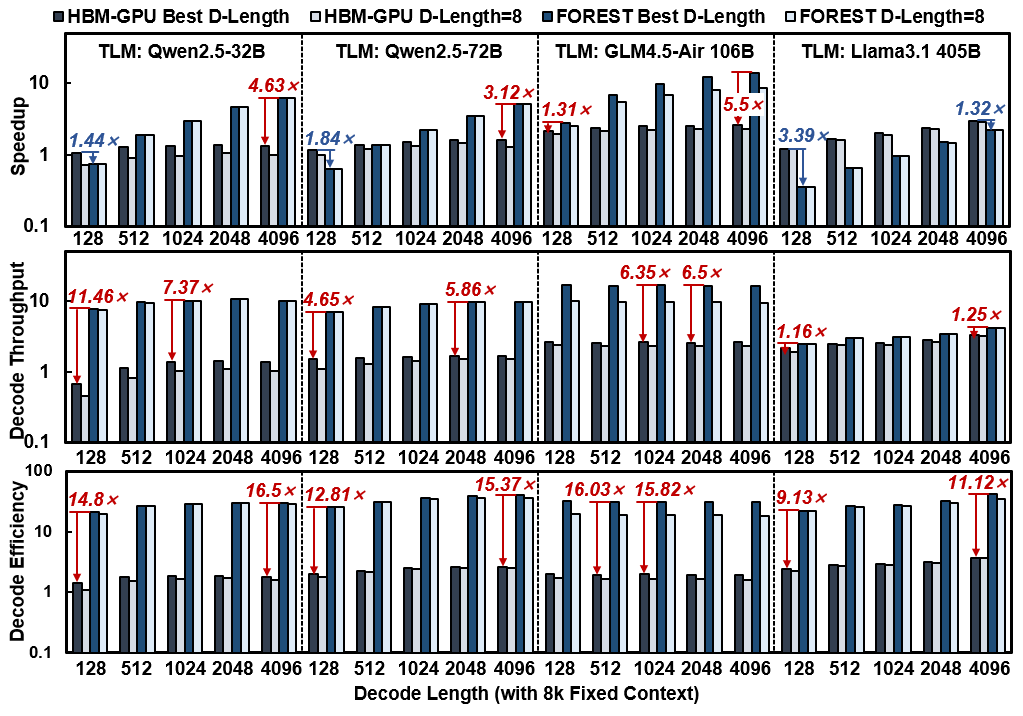}
    \caption{End-to-end performance of FOREST versus HBM-GPU: (top) speedup, (middle) decode throughput, and (bottom) decode energy efficiency. Results are normalized to HBM-GPU target-only inference, with optimal or fixed ($\gamma=8$) draft lengths.}
    \label{fig:e2e_performance}
\end{figure*}

\begin{table}[t]
\centering
\caption{Projected Area and Power Breakdown of FOREST.}
\label{tab:forest_ppa}

\normalsize
\setlength{\tabcolsep}{2pt}
\renewcommand{\arraystretch}{1.02}

\begin{tabular*}{\columnwidth}{@{}p{0.52\columnwidth}@{\extracolsep{\fill}}rr@{}}
\toprule
\textbf{Component} & \textbf{Area (mm$^2$)} & \textbf{Power (W)} \\
\midrule

\multicolumn{3}{l}{\textbf{Level 1: HBF T-Root}} \\
W4A8 DCIM PEs (640) & 26.160 & 2.934 \\
PE-local SRAM (2.5 MiB) & 9.934 & 1.744 \\
Output Converter / Packer & 0.082 & 0.010 \\
Reduction / Control & 2.616 & 0.220 \\
\textbf{HBF Root Subtotal} & \textbf{38.792} & \textbf{4.909} \\

\midrule
\multicolumn{3}{l}{\textbf{Level 2: Dual RRAM D-Roots}} \\
W4A8 DCIM PEs (256) & 5.724 & 0.347 \\
PE-local SRAM (2 MiB) & 3.595 & 0.391 \\
RDR / Gearbox (16 dies) & 1.571 & 0.157 \\
GIU Frame Capture (2 trees) & 0.231 & 1.114 \\
Additional D2D / Refill & 0.451 & 0.090 \\
\textbf{RRAM Logic Subtotal} & \textbf{11.572} & \textbf{2.099} \\

\midrule
\multicolumn{3}{l}{\textbf{Level 3: HBM KV-Root}} \\
K-BFP Logic (512 APUs) & 39.146 & 5.635 \\
V-Link / Local Softmax & 46.632 & 3.316 \\
Global Merge Tree (511) & 20.184 & 1.347 \\
Root Normalizer & 0.474 & 0.063 \\
Additional BFP Control & 1.716 & 0.179 \\
BFP DCIM + Metadata (4.25 MiB) & 16.888 & 5.688 \\
\textbf{HBM Root Subtotal} & \textbf{125.039} & \textbf{16.228} \\

\midrule
\multicolumn{3}{l}{\textbf{Level 4: Memory Stacks}} \\
HBF Stack (512 GB) & 176.64 & 34.680 \\
RRAM Stacks (1 GiB) & 66.96 & 6.221 \\
HBM4 Stack (36 GB) & 140.80 & 37.216 \\

\midrule
\textbf{Compute Roots Subtotal} & \textbf{175.403} & \textbf{23.236} \\
\textbf{Memory Stacks Subtotal} & \textbf{384.40} & \textbf{78.117} \\

\midrule
\textbf{Total Architecture} & \textbf{424.92} & \textbf{101.35} \\
\bottomrule
\end{tabular*}

\end{table}

\subsection{Hardware Implementation}
\label{subsec:hardware_implementation}

Table~\ref{tab:forest_ppa} presents FOREST's area and power breakdown based on 22-nm standard-cell synthesis and memory component modeling. The HBM KV-root dominates the compute-root area, accounting for 125.04~mm$^2$, or 71.3\% of the total. Its K-BFP datapath, V datapath, and 511-node global merge tree occupy 39.15, 46.63, and 20.18~mm$^2$, respectively. Including local storage and supporting circuitry, the HBM root consumes 16.23~W.

The HBF T-root and dual RRAM D-roots occupy 38.79 and 11.57~mm$^2$, consuming 4.91 and 2.10~W, respectively. At the system level, the HBF, RRAM, and HBM memory stacks account for 78.12~W, approximately 77.1\% of FOREST's total estimated power of 101.35~W. The projected three-tree package footprint is approximately 424.92~mm$^2$, accounting for the vertically stacked memory and logic dies.

\subsection{Algorithm-Level Evaluation}
\label{subsec:workload_characterization}

Table~\ref{tab:workload_characterization} summarizes the six target--draft configurations, their W4-equivalent weight footprints, BF16 KV-cache capacities, and GPU-measured acceptance rates. Weight capacities are derived from model sizes, while KV-cache requirements are calculated for 12K-token sequences.

Acceptance rates vary substantially with draft model size and speculation length. For Qwen2.5-32B at $\gamma=8$, increasing the draft model from 0.5B to 3B improves acceptance from 38.49\% to 53.82\%, a gain of 15.33 percentage points. Acceptance generally decreases at larger draft lengths, highlighting the trade-off between candidate generation and effective token acceptance.

To isolate the impact of attention precision, we retain W4-quantized model weights and compare BF16 attention on GPU with a 16-D BFP quantization proxy applied to the target model's Q/K/V tensors. Using Qwen2.5-32B with 0.5B, 1.5B, and 3B drafts at $\gamma=8$, Table~\ref{tab:forest_bfp_sd} reports speculative-decoding behavior across four workloads with 128-token decoding. Average acceptance rates differ by at most 2.13 percentage points, while mean acceptance lengths and verification steps vary by at most 0.141 tokens per step and 3.5 rounds per request, respectively.

\begin{table}[t]
\centering
\caption{Model configurations, memory requirements, and measured acceptance rates.}
\label{tab:workload_characterization}
\small
\setlength{\tabcolsep}{3pt}
\renewcommand{\arraystretch}{1.14}
\begin{tabular*}{\columnwidth}{@{\extracolsep{\fill}}llrrr@{}}
\toprule
\textbf{Target (GPUs)} & \textbf{Draft} &
\shortstack{\textbf{W T/D}\\\textbf{(GB)}} &
\shortstack{\textbf{KV T/D}\\\textbf{(GiB)}} &
\shortstack{\textbf{Acceptance$^{*}$} (\%)\\$\gamma=4/8/16$} \\
\midrule
\multirow{3}{*}{Qwen2.5-32B (1)}
 & 0.5B & 16.25/0.25 & 3.00/0.14 & 55.93/38.49/23.21 \\
 & 1.5B & 16.25/0.77 & 3.00/0.33 & 61.12/44.26/32.68 \\
 & 3B & 16.25/1.55 & 3.00/0.42 & 69.83/53.82/35.76 \\
\midrule
Qwen2.5-72B (1)
 & 1.5B & 36.35/0.77 & 3.75/0.33 & 72.59/57.47/39.64 \\
\midrule
\shortstack[l]{GLM-4.5-Air\\106B (2)}
 & MTP & 53.0/-- & 2.16/-- & 47.63/23.68/12.24 \\
\midrule
\shortstack[l]{Llama 3.1-405B\\(4)}
 & 3.2-1B & 202.50/0.62 & 5.91/0.38 & 73.26/56.67/41.80 \\
\bottomrule
\end{tabular*}
\vspace{2pt}
\parbox{\columnwidth}{\footnotesize\raggedright
$^{*}$ Measured at 4096 output tokens. The 3B draft requires RRAM capacity scaling.
}
\end{table}

\begin{table}[t]
\centering
\caption{SD sensitivity to HBM-GPU/FOREST attention precision.}
\label{tab:forest_bfp_sd}

\normalsize
\setlength{\tabcolsep}{2.5pt}
\renewcommand{\arraystretch}{1.12}

\begin{tabular*}{\columnwidth}{@{}l@{\extracolsep{\fill}}cccc@{}}
\toprule
& \multicolumn{2}{c}{Acceptance (\%)}
& \multicolumn{2}{c}{Verification} \\
\cmidrule(lr){2-3}\cmidrule(l){4-5}
Workload & Avg. & Best & Mean AL & Steps \\
\midrule
General   & 21.22/23.35 & 23.39/26.83 & 2.680/2.821 & 49.0/45.5 \\
Code      & 29.13/28.87 & 34.96/35.14 & 3.252/3.224 & 39.7/40.2 \\
Reasoning & 24.64/25.80 & 31.00/32.81 & 2.894/3.000 & 45.7/44.0 \\
Dialogue  & 30.70/30.70 & 35.26/35.26 & 3.383/3.383 & 37.8/37.8 \\
\bottomrule
\end{tabular*}

\end{table}

\subsection{End-to-End Performance}
\label{subsec:end_to_end_performance}

\begin{figure}[t]
    \centering
    \includegraphics[width=0.86\columnwidth]{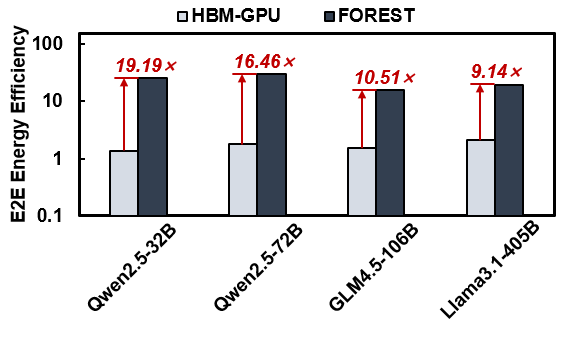}
    \caption{Average end-to-end energy efficiency of FOREST and HBM-GPU at $\gamma=8$.}
    \label{fig:e2e_energy_efficiency}
\end{figure}

Fig.~\ref{fig:e2e_performance} compares FOREST with A800 GPU-based SD across four representative target-model configurations under approximately 8K-token input contexts. FOREST generally achieves greater end-to-end speedups with longer generations, as the increasing contribution of decoding allows its specialized memory--compute pipelines to amortize prefill overheads. At 4096 output tokens, FOREST achieves a 5.5$\times$ speedup over GPU SD for GLM-4.5-Air, whereas GPU SD remains 1.32$\times$ faster for Llama 3.1-405B due to compute-intensive prefill. This contrast highlights the shifting performance bottleneck across model scales: while specialized memory--compute execution benefits decoding, larger models demand greater T-root compute capacity to sustain efficient prefill. Across the evaluated configurations, FOREST further delivers up to 11.46$\times$ higher decode throughput and 16.5$\times$ higher decode energy efficiency over GPU SD.

Fig.~\ref{fig:e2e_energy_efficiency} reports average end-to-end energy efficiency at $\gamma=8$, normalized to GPU target-only inference. FOREST achieves 19.19$\times$, 16.46$\times$, 10.51$\times$, and 9.14$\times$ higher energy efficiency over GPU SD for Qwen2.5-32B, Qwen2.5-72B, GLM-4.5-Air 106B, and Llama 3.1-405B, respectively. Notably, substantial energy-efficiency gains persist even for the 405B workload, despite its lower end-to-end performance, demonstrating that energy and latency benefits do not necessarily scale together.

\begin{figure}[t]
    \centering
    \includegraphics[width=\columnwidth]{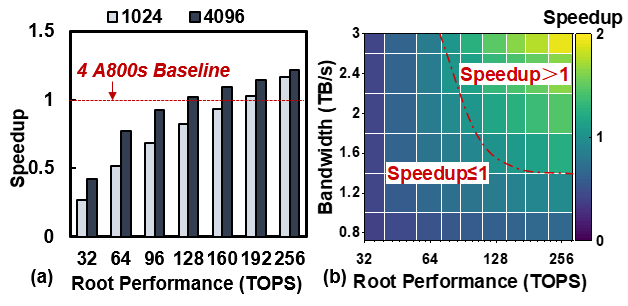}
    \caption{HBF resource sensitivity for Llama 3.1-405B: (a) end-to-end speedup versus T-root compute capacity; (b) joint bandwidth--compute sensitivity. The dashed contour indicates the 1$\times$ break-even boundary.}
    \label{fig:hbf_scaling}
\end{figure}

\begin{figure}[t]
    \centering
    \includegraphics[width=\columnwidth]{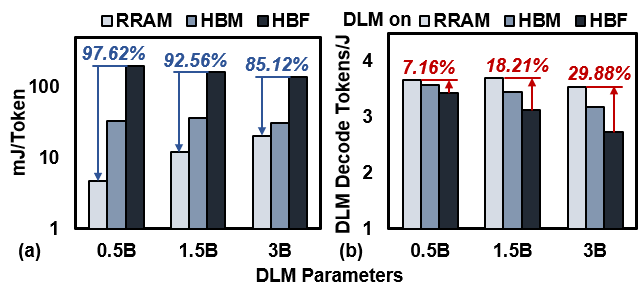}
    \caption{Impact of DLM placement across RRAM, HBM, and HBF: (a) per-token weight-access energy; (b) system-level decode energy efficiency with Qwen2.5-32B as the TLM.}
    \label{fig:rram_role}
\end{figure}

\subsection{Architectural Ablation and Sensitivity}
\label{subsec:architectural_ablation}

The 405B workload exposes a compute-intensive prefill bottleneck, particularly for shorter generations where prefill accounts for a larger fraction of request latency. We therefore examine the impact of HBF bandwidth and T-root compute capacity on end-to-end performance.

Fig.~\ref{fig:hbf_scaling}(a) shows that increasing T-root compute capacity to 128 TOPS enables FOREST to outperform the four-A800 baseline at 4096 output tokens, whereas approximately 192 TOPS is required at 1024 tokens. The joint bandwidth--compute sweep in Fig.~\ref{fig:hbf_scaling}(b) further identifies the resource combinations needed to cross the 1$\times$ performance boundary. Beyond sufficient compute provisioning, HBF bandwidth becomes increasingly important in sustaining performance, limiting the benefits of further compute scaling.

\begin{figure}[t]
    \centering
    \includegraphics[width=\columnwidth]{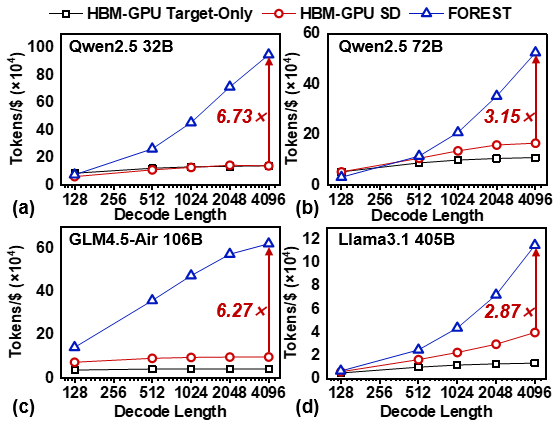}
    \caption{TCO-based cost efficiency (tokens/\$) of HBM-GPU target-only, HBM-GPU SD, and FOREST across output lengths. }
    \label{fig:cost_efficiency}
\end{figure}

\begin{figure}[t]
    \centering
    \includegraphics[width=\columnwidth]{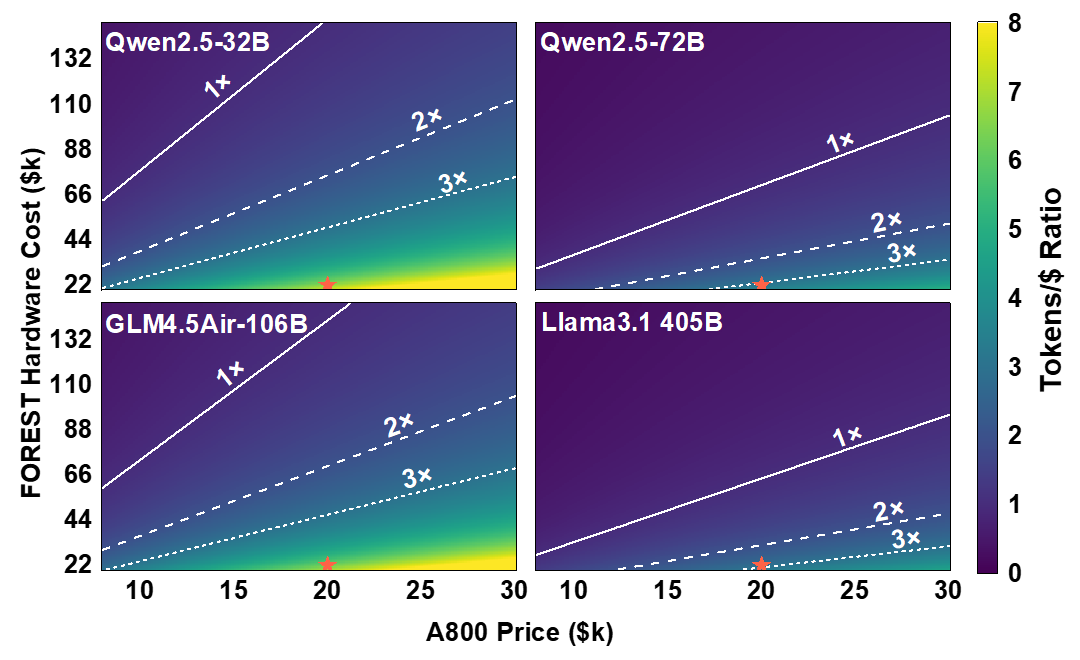}
    \caption{Cost-efficiency sensitivity to A800 and FOREST hardware prices. Colors indicate tokens/\$ gains over HBM-GPU SD, with contours marking 1$\times$, 2$\times$, and 3$\times$ improvements.}
    \label{fig:cost_sensitivity}
\end{figure}

Fig.~\ref{fig:rram_role} evaluates the impact of DLM weight placement across RRAM, HBM, and HBF, with Qwen2.5-32B as the target model. For the 3B draft, RRAM capacity is scaled beyond the nominal 1-GiB configuration to accommodate its weights. As shown in Fig.~\ref{fig:rram_role}(a), RRAM reduces DLM weight-access energy by 97.62\%, 92.56\%, and 85.12\% relative to HBF placement for 0.5B, 1.5B, and 3B drafts, respectively. At the system level, Fig.~\ref{fig:rram_role}(b) reports corresponding decode energy-efficiency improvements of 7.16\%, 18.21\%, and 29.88\% over HBF placement.

\subsection{Cost Efficiency}
\label{subsec:cost_efficiency}

Fig.~\ref{fig:cost_efficiency} compares FOREST with A800 GPU target-only and speculative decoding (SD) baselines using the total cost of ownership (TCO) model described in Section~\ref{subsec:evaluation_methodology}. FOREST's cost-efficiency advantage increases with output length as prefill and fixed execution costs are amortized over more generated tokens. At 4096 output tokens, FOREST achieves 6.73$\times$, 3.15$\times$, 6.27$\times$, and 2.87$\times$ higher tokens per dollar than GPU SD for Qwen2.5-32B, Qwen2.5-72B, GLM-4.5-Air 106B, and Llama 3.1-405B, respectively. Notably, the 405B workload retains a 2.87$\times$ cost-efficiency advantage even when GPU SD achieves higher execution throughput, highlighting the distinction between performance and token-generation cost.

Fig.~\ref{fig:cost_sensitivity} examines the impact of A800 purchase prices and FOREST hardware costs, with contours marking 1$\times$ break-even and 2$\times$/3$\times$ cost-efficiency gains. FOREST maintains a cost-efficiency advantage across a wider range of pricing configurations for Qwen2.5-32B and GLM-4.5-Air, whereas Qwen2.5-72B and Llama 3.1-405B exhibit greater sensitivity to hardware prices.

\section{Related Work}
\label{sec:related_work}

\textbf{Speculative Decoding Accelerators.} SpecPIM~\cite{10.1145/3620666.3651352} explores PIM resource allocation and dataflow co-design for speculative decoding (SD), while LP-Spec~\cite{he2025lpspec} combines LPDDR-PIM with draft-token pruning and dynamic scheduling. 3D-TokSIM~\cite{zhao2025toksim} employs 3D DRAM and token-stationary SRAM CIM, whereas HybridSpec~\cite{huang2026hybridspec} exploits hybrid-bonded memory for heterogeneous SD execution. SADDLE~\cite{wang2026saddle} further introduces adaptive draft lengths and GPU--PIM scheduling. In contrast, FOREST specializes three distinct memory--compute trees for target verification, autoregressive drafting, and KV-cache attention.

\textbf{Memory-Centric LLM Accelerators.} H$^2$-LLM~\cite{li2025h2llm} explores hybrid-bonded DRAM computing, Lincoln~\cite{sun2025lincoln} combines near-Flash computing with speculative decoding, and Lyla~\cite{wang2025lyla} couples RRAM computation with DRAM-resident KV caches. DIAMoND~\cite{liang2026diamond} integrates in-NAND and near-DRAM computing for MoE inference. Recent HBF-based designs include H$^3$~\cite{ha2026h3}, which coordinates HBF and HBM placement, FlashAccel~\cite{wang2026flashaccel}, which optimizes HBF-aware data access and management, and FLINT~\cite{oliveira2026flint}, which improves HBF burst delivery and flash management. FOREST extends heterogeneous memory integration toward SD-specific compute provisioning and coordinated execution across HBF, RRAM, and HBM.

\textbf{In-Memory Attention and Numerical Co-Design.} FlashAttention~\cite{dao2022flashattention} reduces attention memory traffic through tiling and online softmax. AttAcc~\cite{park2024attacc}, CHIME~\cite{liu2026chime}, and SoftmaxPIM~\cite{kang2026softmaxpim} accelerate attention through near-memory processing and dataflow optimization. P$^3$-LLM~\cite{chen2026p3llm} explores hybrid numerical formats and fused low-precision computation, while Harmonia~\cite{wang2026harmonia} extends block-floating-point (BFP) processing to attention and KV caches. FOREST instead coordinates producer-side block-scaled Q/K/V generation with BFP attention computation in a separate HBM-resident DCIM tree.

\section{Conclusion}
\label{sec:conclusion}

This paper presented FOREST, a heterogeneous memory-centric architecture that coordinates HBF, 3D-stacked RRAM, and HBM memory--compute trees for efficient speculative decoding. FOREST combines compute-aligned RRAM weight delivery with cross-tree numerical co-design, enabling latency-sensitive drafting and block-floating-point (BFP) attention processing through specialized compute paths. Evaluated across six speculative decoding configurations spanning 32B--405B target models, FOREST achieves up to 5.5$\times$ end-to-end speedup, 11.46$\times$ higher decode throughput, and 16.5$\times$ higher decode energy efficiency over A800 GPU-based speculative decoding baselines. Under a modeled total cost of ownership, FOREST further delivers up to 6.73$\times$ higher token-generation cost efficiency. Our sensitivity analysis reveals that balancing HBF bandwidth and compute provisioning is essential for overcoming prefill bottlenecks and realizing end-to-end performance gains. These findings highlight the potential of coordinated, role-specialized memory--compute architectures to address the diverse resource demands of speculative LLM inference.


\bibliographystyle{ACM-Reference-Format}
\bibliography{refs}
\end{document}